# A Photonic Crystal Receiver for Rydberg Atom-Based Sensing

*Hadi Amarloo, Mohammad Noaman, Su-Peng Yu, Donald Booth, Somayeh Mirzaee, Rajesh Pandiyan, Florian Christaller and James P. Shaffer, Quantum Valley Ideas Laboratories, 485 Wes Graham Way, Waterloo, Ontario, Canada N2L 0A7*

**Abstract:** Rydberg atom-based sensors use atoms dressed by lasers to detect and measure radio frequency electromagnetic fields. The absorptive properties of the atomic gas, configured as a Rydberg atom-based sensor, change in the presence of a radio frequency electromagnetic field. While these sensors are reasonably sensitive, the best conventional radio frequency sensors still outperform Rydberg atom-based sensors with respect to sensitivity. One approach to increase the sensitivity of Rydberg atom-based sensors is to engineer the vapor cell that contains the atomic gas. In this work, we introduce a passive, all-dielectric amplifier integrated into a Rydberg atom-based sensor vapor cell. The vapor cell is a combination of a slot waveguide and a photonic crystal. The structural features of the vapor cell yield a power amplification of ~24 dB. The radio frequency electromagnetic field is coupled adiabatically into the slot waveguide and slowed to increase the interaction between the radio frequency field and the atoms to effectively amplify the incoming signal, i.e., increase the Rabi frequency on the radio frequency transition. The work shows the utility of vapor cell engineering for atom-based quantum technologies and paves the way for other such devices.

A receiver is a device that converts electromagnetic radiation to a usable form. Rydberg atom-based receivers use atoms to convert free-space, radio frequency (RF) electromagnetic waves into an optical signal that can be read out on a photodetector at baseband[1–4]. The photodetector converts the optical signal into an electrical signal. Multiple probe frequencies can be applied simultaneously so that the probe laser can be readout across a range of frequencies at the same time[5]. Rydberg atom-based receivers, particularly those that are all-optically read out, have a number of advantages that conventional antenna-based receivers do not possess. The receiver is more electromagnetically transparent, enabling the mounting of multiple receivers on a platform of limited size; self-calibrated, which allows amplitude to be meaningfully used in applications like radar; and read-out at baseband, eliminating the need for mixers and RF amplifiers. Because Rydberg atom-based receivers are coherent, they can sense below the thermal noise limit. Thermal noise affects them through the modification of the decay rates of the Rydberg states, so-called black body radiation[6]. Rydberg atom-based sensors are frequently limited by shot noise present in the probe laser or projection noise in the sensor, i.e., atomic shot noise[2,7]. There is no known way to overcome the shot noise present in the classical laser fields and uncorrelated atomic sample without using additional quantum resources, i.e., entanglement or squeezed light, which are impractical at this time and radically change the potential to realize small size, weight and power (SWaP) systems given the current state of the field[8]. Although Rydberg atom-based sensors have attractive properties that potentially make them superior to conventional receivers in ways that leverage their unique features, they have not been demonstrated to be as sensitive.

A path forward to improve the sensitivity as well as other properties, like RF transparency, for a broad range of transformative communications and radar applications is to engineer the vapor cell to enhance the signal[9]. To improve the sensitivity of Rydberg atom-based receivers, we engineered an all-dielectric vapor cell so that it comprises a passive amplifier. The all-dielectric vapor cell is constructed from glass and silicon. The methods we have developed for building the vapor cell are suitable for scaling to industrial production. The receiver is based on photonic crystal and waveguide ideas at RF[10–12]. To make an amplifier, we enhance the electric field at the position of the atoms, which serve to transduce the RF signal to an optical signal, so that we can increase sensitivity. We use a monolithic photonic crystal vapor cell to accomplish this task by slowing the radio frequency wave and concentrating it in a slot waveguide[13–15]. The signal levels in a bare vapor cell are limited because ultimately the interaction of the atom with the incident electromagnetic field is determined by the absorption rate times the coherence time, i.e. the coupling constant between the atom and the field and how fast the atom decays. Photonic crystal cavities and structures are ideal for integrating with alkali vapor cells, the key element for Rydberg atom-based sensors, because the wavelengths of the RF radiation have the same size scale as the vapor cell and its features. Photonic crystal cavities and structures can be used to engineer the local density of states to improve the interaction of the atoms with the RF field. We refer to the device as a photonic crystal receiver (PCR).

The PCR system consists of the photonic crystal vapor cell, an optical circuit to connect the laser light to the PCR, and the laser, detection and digital control systems, including the signal processing. The prototype PCR presented in this paper has a demonstrated power gain of ~24 dB. The general principles demonstrated in the paper can be used to make passive, vapor cell integrated amplifiers with much larger gain. To date, others proposing receivers using Rydberg atom-based sensing have used large, blown glass vapor cells which are not competitive with modern receivers. These efforts have resorted to metal constructions to create a slot waveguide, thus eliminating a key feature of the receiver, its electromagnetic transparency[16–18]. Our vapor cell technology allows us to build a receiver that retains its dielectric nature and can have a sensitivity that has the potential to exceed that of conventional receivers.

## Results

**Group velocity and index**

The structure of the PCR is shown in Fig. 1. A detailed discussion of the structure and its fabrication may be found in the methods section. The RF electromagnetic wave's interaction with the atoms depends on the strength of the RF electric field and the interaction time. Increasing either factor, will increase the atomic response to the RF electromagnetic wave. Due to the discontinuity in the refractive index between the free-space slot and silicon frame, the waveguide supports a mode with the RF electric field highly confined in the slot region, shown in Fig. 2c, increasing its strength at the position of the atoms. The photonic crystal is designed to decrease the group velocity, $v_g$, of the RF electromagnetic wave that is coupled to the waveguide mode around a specific RF frequency. If the RF electromagnetic wave has a frequency close to the band-edge of the photonic crystal waveguide, $v_g$ can be reduced by several orders of magnitude, increasing the interaction time between the atoms and the RF electric field. Fig. 2a shows the calculated band structure of the device, with the group index, $n_g = c/v_g$, shown in Fig. 2b.

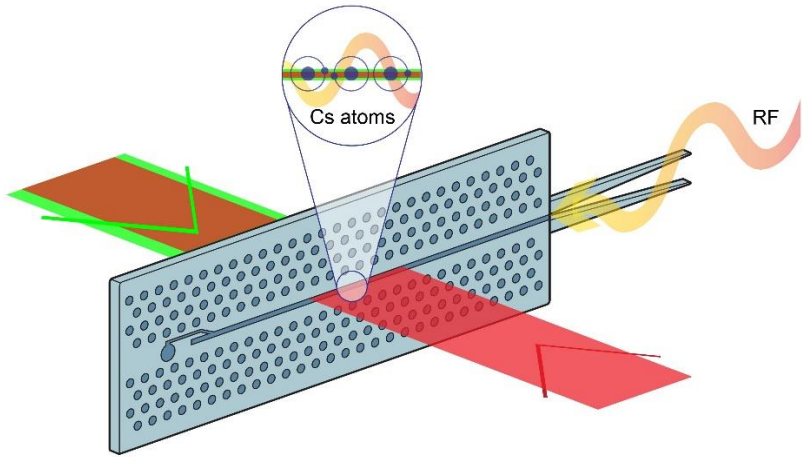

**Figure 1:** Photonic crystal vapor cell with slot region filled with Cs atoms, and taper structure for mode conversion of the incoming free space RF electromagnetic wave. The holes are organized to produce a photonic crystal that slows a RF electromagnetic wave around a specific frequency. The slot is hermetically sealed with glass on both sides and filled with Cs atoms. The device uses a thermally activated getter source to load the Cs atoms, shown as the disc in the circular pocket separated from, but fluidly coupled to the slot.

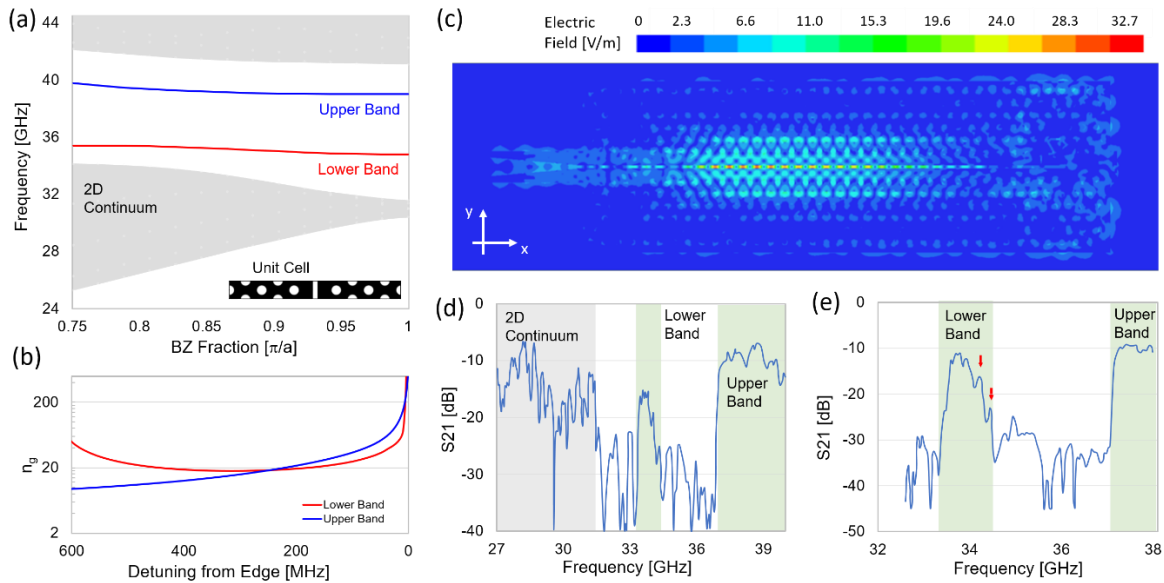

**Figure 2:** (a) Band structure of the one-dimensional guided modes and (b) the group indices, $n_g$, near the band edge. The holes in the unit cell have a diameter of 1 mm. Adjacent holes are separated by 2 mm (c) Calculated field profile of a one-port device. (e) Measured transmission spectra of a two-port device and (f) zoom-in data set showing band-edge resonances (red arrows).

The group velocity mismatch induced by reflection from discontinuities along the one-dimensional structure of the PCR can be partially mitigated by adiabatic tapering. The group velocity is ramped gradually instead of abruptly. The tapering is advantageous as it facilitates the maximum RF input entering the one-dimensional defect where the atoms are located. In the spectral domain, the tapers reduce sharp resonance features. In our device, tapers of 18 mm, comparable to the length of the main photonic crystal section, are fabricated at the input of the device. We observed significant suppression of Fabry-Perot-like resonances using the tapers, but the reflections are not entirely eliminated. A lower frequency band of the photonic waveguide shows stronger resonance features, as the tapers are not optimized for this band. The suppression of the Fabry-Perot resonances at the device resonance in comparison to the features associated with the lower frequency band verifies that the tapers work, but longer tapers can lead to significant improvements. The input RF coupling, facilitated by the tapers, to the current PCR is ~10%.

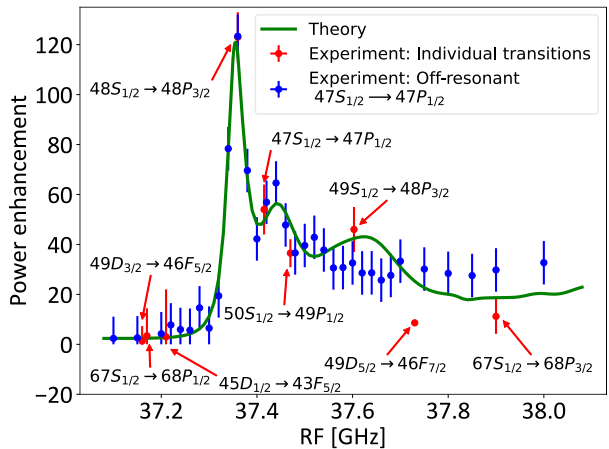

Figure 3: Calculation and measured data for the sensitivity enhancement of the PCR. The calculation is shown by the green line. The red points show the result of a set of measurements resonant with atomic transitions (labeled on the figure). The red points show the result of off-resonant measurements taken on the $47S_{1/2} \rightarrow 47P_{1/2}$ transition (resonant at 37.41 GHz). Both sets of measurements are taken using 20 mm x 0.5 mm Rydberg laser beams, covering the active region of the device. The large resonance is due to the slow light effect while the enhancement at frequencies greater that ~37.8 GHz is associated with the enhancement of the slot waveguide.

The response of the PCR is calculated using an electromagnetics calculation (Ansys HFSS) to determine the expected power enhancement as a function of RF frequency, Fig. 3. The enhancement is determined by calculating the Purcell factor. A dipole is placed in the center of the slot and the power emitted by the device when it is excited by the dipole is compared to the same dipole in free-space. The loss associated with the tapers is taken into account by the method. The calculation is shown in Fig. 3. The enhancement factor at the peak is around 124 (21 dB) at 37.36 GHz, the resonance frequency. Below the resonance frequency, the enhancement drops rapidly. The rapid drop in the response is due to the fact that these frequencies fall within the band gap of the device. For the frequencies above the resonance, there is a smaller enhancement which is a consequence of the electric field confinement in the slot. The electric field confinement provides enhancement over a wider range of frequencies than the slow wave effect. The shape of the response curve shows that both the slow light and confinement effects are working to enhance the response to the incident RF electromagnetic wave.

**Gain calculations and measurements**

The band structure of the PCR was obtained by measuring the transmission of a RF electromagnetic wave through the device as a function of frequency. A RF vector network analyzer (VNA) is used to excite the guided mode and measure its transmission. In order to couple the RF excitation into the device, we fabricate adiabatic silicon tapers that efficiently connect to conventional metal RF waveguides. The VNA performs a frequency sweep of the RF wave and records the transmitted power. Fig. 2d-e shows the transmission through the device. We observe the two predicted guided bands, within the bandgap where transmission sharply drops. The higher frequency band above 37 GHz is the band of interest for this work.

The slow wave group velocity onset near the bandgap creates resonances at the edge of the stop band. As the frequency approaches the stop band, $v_g$ decreases and becomes mismatched to the waveguide in the photonic crystal. The impedance mismatch leads to reflections at the two waveguide-taper interfaces. The interfaces form a Fabry-Perot type resonator, leading to the observed resonances. Fig. 2e shows a zoomed-in spectrum of the two guided bands. The resonances manifest as transmission peaks visible on the lower band. The resonances near the upper band are suppressed due to the impedance matching structures. Due to constraints on the size of a device that can be built on a four-inch silicon wafer, the impedance matching of the PCR is not ideal.

Since the device is operated near the sharp onset of slow-light, near the bandgap to maximize field enhancement, we developed a method to tune the narrow device resonance to an atomic state. The band structure can be uniformly tuned by utilizing the evanescent field. The thickness of the photonic crystal receiver is smaller than the RF wavelength of the device resonance. The guided mode consists of evanescent fields extending beyond the top and bottom surfaces of the device. By inserting a sheet of dielectric, e.g. glass, into the evanescent field, it is possible to tune the bands toward lower frequency. We vary the distance between the dielectric sheet and the device surface to tune the device resonance to that of the atoms (see *Supplementary* Fig. S2).

To test the enhancement of RF field inside the PCR, we measured the response of the atoms in the device. An RF electromagnetic plane wave near the resonance frequency of the device was applied to the input port of the PCR and Rydberg EIT measurements were performed. Counter-propagating probe and coupling laser beams are shaped to fill the channel over a length of 20 mm. We chose to make measurements for Rydberg states that match the range of RF frequencies located in the gain band of the PCR. The enhancement was referenced to experiments in a conventional vapor cell placed at the same position as the input to the PCR. Figure 3 shows a result of the measured gain overlaid on a full electromagnetics calculation of the device response using HFSS. Off-resonant measurements were performed for the $47S_{1/2} \rightarrow 47P_{1/2}$ transition to obtain a larger number of data points. Both the on- and off-resonant measurements show good agreement with the electromagnetics calculations, suggesting that the calculations accurately describe the behavior of the PCR near the band gap edge, in the center of the device. Both the slow light and confinement enhancement effects are playing a role in the amplification of the detected RF electric field because a clear resonance is found near the band edge indicating the slow-wave effect and a broader gain band is present associated with the slot waveguide effect.

The RF electric field along the PCR axis was investigated by focusing the counter propagating probe and coupling laser beams to a spot along the channel. The spot size is 200 μm. The PCR is translated along the slot-axis with 30 μm precision to measure the response. Figure 4a shows the variation in the EIT splitting as the position along the slot is varied while the PCR is illuminated by the RF electromagnetic wave. The PCR is designed with a periodic lattice of 2 mm which governs the local RF field inside the channel. The spatial RF field induced splitting of the EIT transmission measurements show the 2 mm RF field modulation period. The RF field is incident from the left side of the plots in Fig. 4a. The position $Z = 0$ marks the input end of the PCR. The RF field dies out at increasing distance from the input due to imperfections in the machining of the holes of the photonic crystal. A simulation with the randomized position error agrees well with the measured RF field as a function of length.

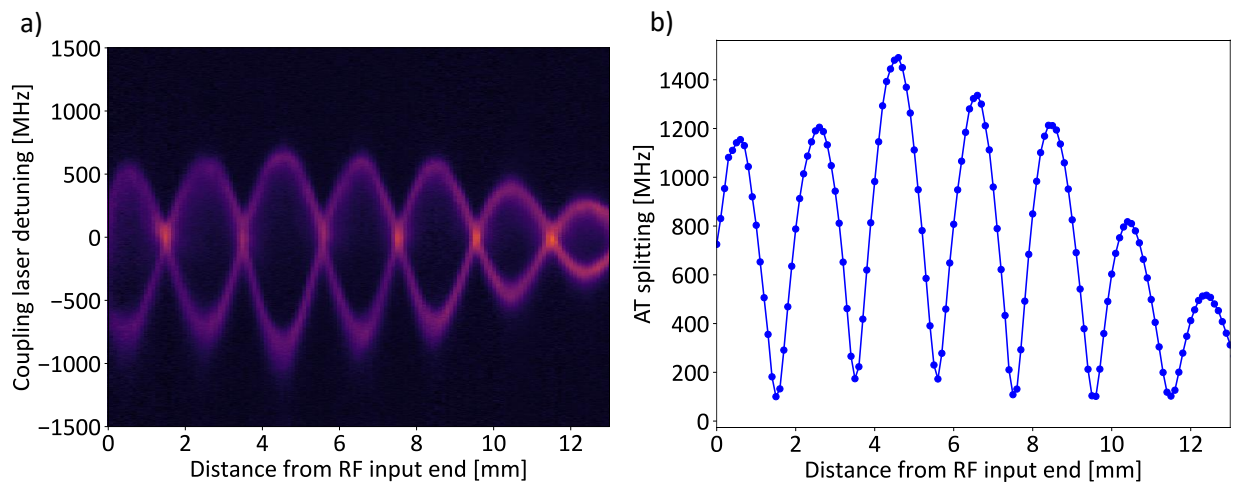

Fig. 4. The Autler-Townes splitting as a function of position along the slot channel in the PCR.

The amplification of the RF electric field by the PCR was measured by comparing the Autler-Townes splitting in the device to a reference frequency splitting obtained using the same commercial vapor cell as the prior gain measurements. Both the PCR and conventional vapor cell were measured at the same distance from a horn antenna under RF plane wave illumination. Figure 4 shows the Autler-Townes splitting at the input end of the PCR. Figure 5a shows the Autler-Townes splitting in the conventional vapor cell as a function of RF detuning, while Fig. 5b shows the Autler-Townes splitting in the PCR under the same conditions. The PCR clearly shows an enhancement in the RF field which results in the larger frequency splitting.

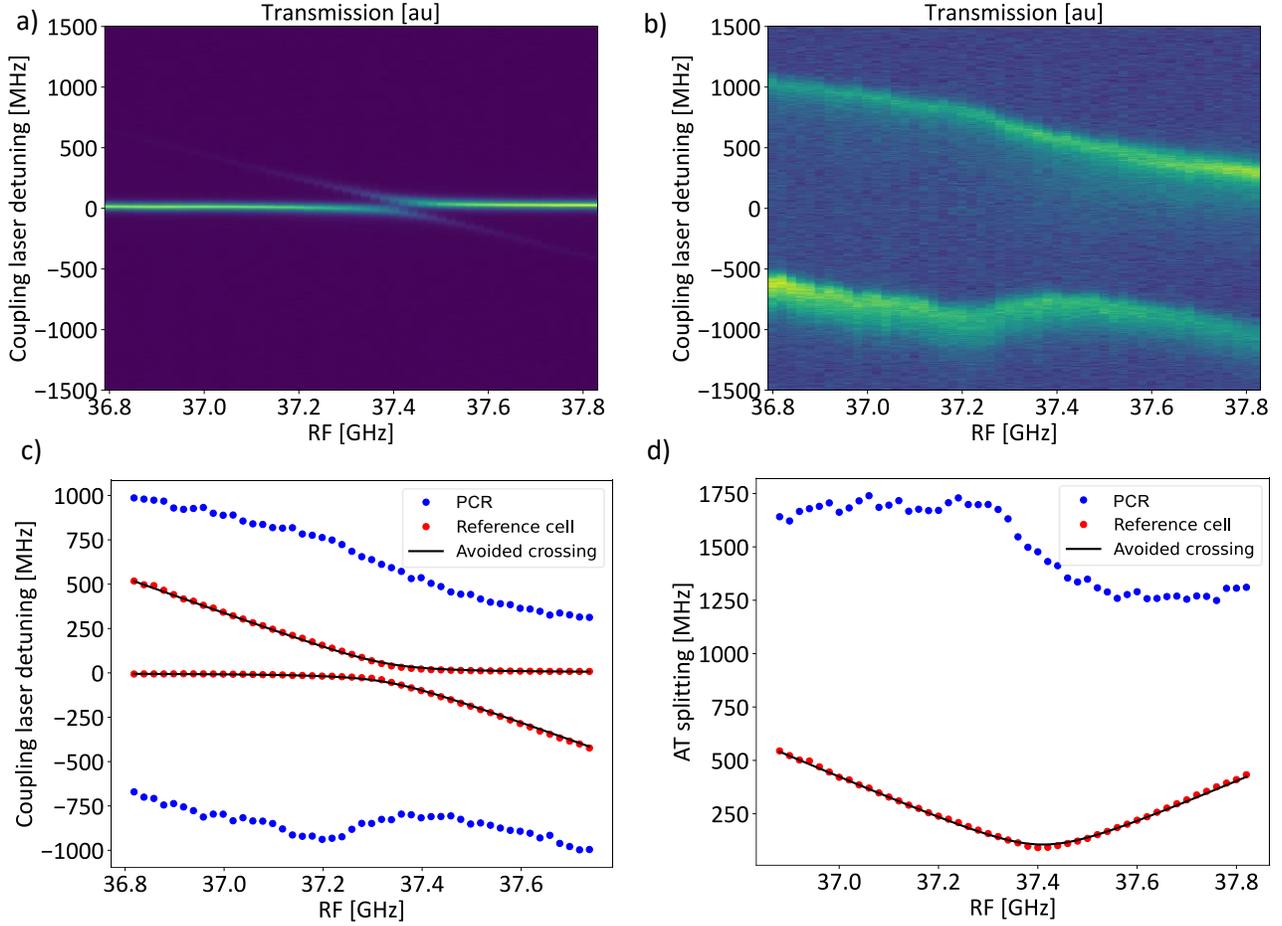

Figure 5: The Autler-Townes splitting of the $47S_{1/2} \rightarrow 47P_{1/2}$ transition as a function of RF frequency for fixed RF power. a) Splitting in a reference vapor cell. b) Splitting in the PCR at z=4 mm. c) Fitted peak centers for each Autler-Townes peak in both (a) and (b). d) The Autler-Townes splitting for (a) and (b). A fit is shown (black line) for the reference cell data to an avoided crossing model. The ratio of the PCR splitting to the avoided crossing fit gives the gain of the PCR.

The amplification of the RF electric field due to the slow light effect in the photonic crystal results in a stronger coupling between the atom and the field that is dependent on the frequency of the photon, similar to a polariton state. The highly dispersive behavior near the resonance can be compared to the behavior of an atom in a cavity[19]. Off resonant measurements provide a way to estimate a continuous power gain for the entire frequency band. The measurement is performed by varying the detuning of the RF field compared to a particular Rydberg transition. In the reference cell, the Autler-Townes splitting $\Delta \nu_{AT,\delta}$ is

$$\Delta \nu_{AT,\delta} = E_+ - E_-,$$
$$E_\pm = -\frac{\delta}{2} \pm \frac{1}{2}\sqrt{\Omega^2 + \delta^2},$$

where $\Omega$ is the splitting at resonance of the RF transition, and $\delta$ is the detuning of the transition from the resonance. The form for $\Delta \nu_{AT,\delta}$ results in an avoided crossing in the spectrum, as seen in Fig. 5c. By comparing the frequency splitting of the peaks in a reference cell and in the PCR, as shown in Fig. 5d, it is possible to extract the estimated power gain as a function of the RF frequency. However, the data in Fig. 5 is complicated by the energetically nearby $47P_{3/2}$ Rydberg state. We implemented a three-level model to interpret the results. The details of the model can be found in the supplemental section of the paper.

The estimated power gain from the three-level model is plotted in Fig. 6. The RF power enhancement is observed to be around 270 (24 dB). A stronger driving field will lead to a dressing of the ground state which is visible as a splitting of the energy level at zero detuning. The $47P_{3/2}$ state a few GHz away will have an influence on the Autler-Townes states around zero detuning for a strong driving field. The effect of the $47P_{3/2}$ state is seen in the kink in the data in Figs. 5b and 5c.

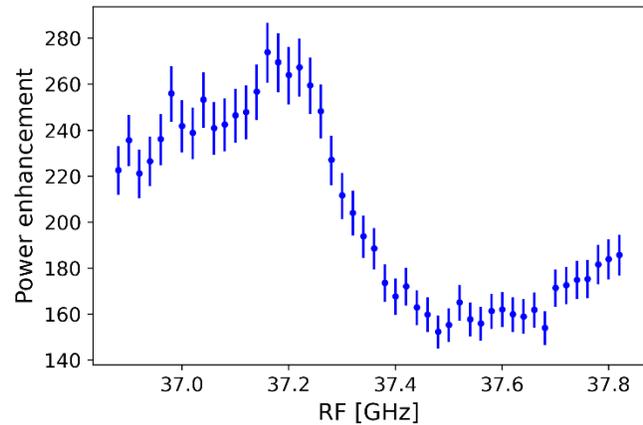

**Figure 6:** RF power enhancement estimated by comparing the PCR splitting with respect to a commercial vapor cell. The measurement is performed using a small volume of 100 um diameter at the maximum splitting site in Fig. 4.

The results shown in Fig. 3 are different from Fig. 6 because the data was taken in different parts of the PCR. As shown in Fig. 6, for the frequency in the bandgap, the enhancement increases, which is expected due to the cavity formed between the input coupler and the photonic crystal. This effect is discussed further in the supplementary section of the paper. For lower RF frequencies, the PCR response will be less sensitive to fabrication tolerances because the wavelength is longer when compared to the fixed manufacturing error. The results in Fig. 6 show that cavities can be implemented in PCRs to further increase their gain.

We also used the PCR to detect fast RF pulses because time-dependent signals are the most likely for applications like radar and communications. The basic scheme is the same as that described in Ref. [20]. The probe and coupling lasers are locked on resonance and a fast ac-coupled photodiode (50 MHz) is used to measure the probe transmission signal. The signal is fed into an FPGA where it is digitized and passed through a matched filter. 10 µs RF pulses are introduced to the PCR. The matched filter signal output is plotted in Fig. 7. The detected RF pulse signal is distinct from the background noise and can be used as trigger in real time data acquisition. The SNR of the detected signal is 11 which corresponds to a jitter of 0.78 µS in the trigger time. The incident RF field is estimated to $6.8 \pm 0.2$ mV/cm.

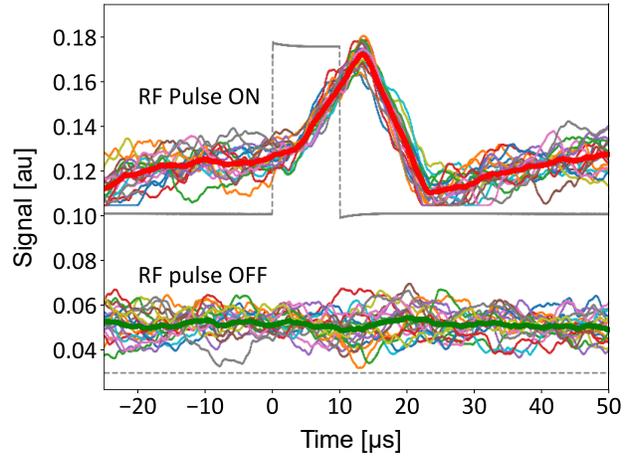

Fig. 7: A comparison of the time-dependent matched filter response in the EIT signal with and without the presence of an RF pulse. A peak clearly appears in the matched filter signal when the RF pulse is present. For clarity, a vertical offset is added to the RF on data.

## Discussion

We have shown that it is possible to construct a passive amplifier for a Rydberg atom-based RF receiver that is built into the vapor cell. The amplifier is entirely dielectric and amplifies the RF power by ~24 dB. The input was designed for a Gaussian beam so a dish or focussing element can add additional gain. The vapor cell uses waveguiding and photonic crystal design features to increase the RF electric field strength and slow the RF electromagnetic wave so as to increase the interaction between the atoms and RF electric field. The experimental results show that the system can be viewed as a cavity for the atoms, effectively increasing the Purcell factor, i.e. the local density of states. Clear avoided crossings between the device resonances and the atoms are observed. The structure is a standing wave cavity so the amplification is periodic in the device. A matched filter was used to detect pulses applied to the device to show that a time dependent signal can be detected. The gain can be further increased by adding a cavity to the device, as demonstrated in the non-ideal impedance matching of the current realization. Travelling wave structures are also possible. In the future, it will be possible to increase the gain by several orders of magnitude, especially at lower frequencies where manufacturing tolerances are more forgiving, since the wavelength increases.

## Acknowledgements

This work has been supported by Defence Research and Development Canada (DRDC) under the "Innovation for Defence Excellence and Security" IDEaS program "Quantum Leap: Shrinking sensing technologies for field operation" (contract number: W7714-217517/001/SV1), FedDev

## Methods:

### Overview of Rydberg atom-based Radio Frequency Sensors

Rydberg states are highly excited states of an atom or molecule characterized by large principal quantum numbers, $n$, and long lifetimes. It has long been understood that the large polarizability and strong dipole transitions between Rydberg states make them highly sensitive to electric fields [21]. The RF electric field measurements described in this paper rely on near-resonant transitions and the associated large transition dipole moments between neighboring, or nearby, Rydberg states that scale as $n^2$, for example, $\mu \approx 1749$ $ea_0$ for transitions to neighboring levels for $n \sim 65$ in $^{87}$Rb. Here, $e$ is the elementary charge and $a_0$ is the Bohr radius.

The RF electric field coupling between two close-lying Rydberg states,

$$\Omega_{RF} = (\mathbf{\mu} \cdot \mathbf{E})/\hbar, \qquad (1)$$

can be large even when the RF electric field amplitude, **E**, is weak. In our approach, the coupling to a RF electric field, $\Omega_{RF}$, will cause an Autler-Townes spectral splitting proportional to $\Omega_{RF}$. Small RF electric field amplitude results in a signal that is straightforward to observe spectroscopically in a room temperature vapor cell with modern frequency stabilized diode lasers, as long as one can utilize a sub-Doppler method for detecting the Autler-Townes splitting of the transition. The higher the spectral resolution of the sub-Doppler method, the better the sensitivity can be made, because a smaller frequency difference can be measured[22,23]. Rydberg atom electromagnetically induced transparency (EIT) and electromagnetically induced absorption (EIA) are suitable tools[1,2,22]. The approach is a self-calibrated method for measuring RF electromagnetic fields, because it is linked to precision measurements of the of the atom, $\mu$, and physical constants. We take advantage of the many resonances between Rydberg states which span from MHz-THz to cover a quasi-continuous spectral range. Continuous coverage over a large range, ~100 GHz, can be achieved using off-resonant sensing, albeit with an associated loss of sensitivity. The RF electric field can still be sensed when the EIT/EIA spectra is not split. The amplitude of the EIT/EIA signal can be used to determine the RF field strength[6]. The amplitude change can be used for time-dependent signals to obtain the highest sensitivity[20]. In the amplitude sensing regime, the signal is not self-calibrated, since the amplitude of the EIT/EIA signal also depends on the Rabi frequencies of the laser fields.

**Design of the Photonic Crystal Receiver**

The central part of the PCR has a photonic crystal structure which is a triangular lattice of holes in the silicon substrate. A photonic crystal waveguide is made by removing one row of the holes comprising a one-dimensional defect in the photonic crystal structure. A slot region is implemented along the center line of the defect to contain the atoms. Cladding layers of glass are bonded on each side of the silicon substrate to hermetically seal the atoms inside the vapor cell. The triangular lattice of holes creates a two-dimensional bandgap designed for a frequency range of interest. The combination of the slot and photonic crystal creates a one-dimensional defect mode because RF scattering into the two-dimensional photonic crystal lattice is forbidden in the bandgap. By choosing the geometrical parameters of the one-dimensional defect and lattice of holes, the device is engineered to create the desired RF electric field enhancement and slow-wave properties.

In order to achieve efficient coupling of the free space propagating RF electromagnetic wave to the slow electromagnetic wave RF mode of the waveguide, several mode converters are required. The mode converters are designed to couple a Gaussian beam into the PCR. There is an adiabatic taper at one side of the device that acts to convert the incoming RF electromagnetic wave from free space into the device mode. In addition, slowing the RF electromagnetic wave is done gradually over ten-unit cells of the waveguide to gradually slow the incoming RF wave and reduce reflections from the group velocity mismatch.

The response of the PCR was calculated using a full-wave numerical simulation tool (Ansys HFSS). The unit cell of the photonic crystal waveguide is shown in Fig. 2a. We designed the PCR for a resonant frequency of 37.4 GHz. The lattice constant of the photonic crystal is 2 mm. The silicon slab thickness is 1.5 mm. The hole diameter is 1 mm and the width of the slot is 0.5 mm. Fig. 2c shows the calculation of the electric field distribution in the device. As shown in Fig. 2c, the electric field is highly confined in the slot region, where the atoms are located. The increase in the RF electric field from the confinement is a factor of 30 times, which is close to what is expected according to the boundary conditions, where the normal component of the electric field is scaled inside the slot by the square of the refractive index of silicon.

The MIT Photonic Bands (MPB) package is used to calculate the band structure of the device, from which an estimate of the group index, $n_g$, is found. We calculated the band structure for the two-dimensional photonic crystal and the one-dimensional defect structure unit cells. Fig. 2a shows the band structure of the one-dimensional guided mode. We identify two guided bands within the bandgap, where propagation in the two-dimensional photonic crystal is forbidden. The band structure calculation enables a comparison between the theory and RF transmission experiments performed with the fabricated device. The one-dimensional guided bands are separated by one-dimensional bandgaps. The onset of the bandgap is where the slow-light effect is predicted. Fig. 2b plots the calculated $n_g = \frac{c}{v_g}$ for the guided modes. We predict significant increase in $n_g$ within 500 MHz of the edge of the bandgap, where the group index is singular.

Group indices in the range of tens to a few hundreds are predicted. We calculated the electric field in the slow-light regime and found a scaling with $1/\sqrt{v_g}$.

The pattern in the silicon substrate (including the photonic crystal structure, slot region, and tapers) is fabricated using laser machining. Two 500 µm-thick borosilicate glass windows are bonded to the double-side polished, high-resistivity silicon substrate surfaces using anodic bonding at a temperature of 400°C and voltage of 1kV. The bonding process provides a strong bond that is leak-tight for high vacuum. The high temperature of the anodic bonding process can result in outgassing that increases the background pressure in the PCR. In order to minimize outgassing, low-temperature anodic bonding (<250°C) is employed to seal the PCR. Before final sealing of the device, a Cs getter pill (SAES) is loaded in the body of the photonic crystal vapor cell. The anodic bond to seal the device is performed in a vacuum chamber at $10^{-6}$ Torr. The getter pill is heated with a 1064 nm laser to release the Cs atoms.

A replica device was fabricated with identical photonic waveguide and tapers. The replica enabled us to build a transmission port on the chip, i.e., we built a two-port device. Instead of the external silicon taper for matching a Gaussian incoming beam to the waveguide mode, we fabricated tapers that are optimized for coupling the guided mode to a conventional WR-12 RF waveguide. We attach two fused silica sheets to the surfaces of slab photonic crystal to simulate the effect of cladding glass. The chip frame, silicon tapers, and silica sheets are cleaned and then assembled.

To test the two-port device, we suspend the glass-clad chip over two thin strips of plastic tape. We then connect the device input and output ports to WR-12 waveguides. The WR-12 waveguides are mounted on three-axis translation stages for alignment. The silicon tapers extending from either port of the device are inserted into the WR-12 waveguide opening. The waveguides are connected to high-frequency coaxial cables and then to the vector network analyzer (VNA). We optimize the VNA spectrum by maximizing RF transmission in the propagation band and minimizing the spurious transmission in the bandgap. We record measurement results from the VNA spectra. In the tuning rate measurements, we fabricated a plastic holder using 3D printing to hold and position the tuning glass plate. An additional translation stage is added to position the glass plate and step its distance to the device.

We obtain information about the propagation bands and bandgap from the VNA spectra, Fig. 2e-f. The RF transmission is inhibited in the bandgap but can propagate in the transmission bands. We gauge the performance of the group velocity matching taper by inspecting the spectrum near the edge of the bandgap. Without the matching taper, the propagation band shows a series of Fabry-Perot-like transmission peaks. The sharpness of these peaks signifies high reflection from mismatch. The spacing of the peaks provides rudimentary information on the group velocity, where the spacing reduces as the frequency approaches the band edge.

Detection of microwave fields in the PCR is done using electromagnetically induced transparency (EIT). The two lasers required for the EIT are a probe laser at 852 nm, and a coupling laser at

509 nm. The probe laser is generated using an external cavity diode laser offset locked to a peak of a ULE Fabry-Perot cavity. The coupling laser is generated using a frequency-doubled external cavity diode laser. The 1018 nm fundamental is scanned over the EIT peak to detect the frequency splitting in the EIT transmission features due to the presence of a RF field.

The probe and coupling lasers counter propagate through the PCR, normal to the plane of the PCR. For the resonance and gain measurements, the beams are 20 mm x 0.5 mm. To measure the slot electric field pattern, the beams are reduced to a waist of 200 μm. The probe beam is detected on an amplified silicon photodiode.

To translate the PCR along with the RF antenna, we mounted both on a movable platform. The laser beams were fixed to maintain their alignment. The configuration ensures the relative distance between the antenna the PCR is held constant. The stage is moved to a different position along the Z axis and thus the laser beams address the desired position along the channel.

In order to verify the glass plate tuning method and characterize the tuning rate, we built a test device with input and output tapers. Both group velocity matching tapers and ordinary group velocity waveguide to conventional metal waveguide tapers were designed and utilized to couple RF waves from the VNA through the device and back to the VNA. The band structure of the device was monitored with the VNA by measuring the RF transmission. The glass tuning plate was held in close vicinity to the device, ~2.2 mm, with a three-dimensional printed plastic holder. The distance between the device and glass plate was controlled using a micrometer stage. The nominal edge of the propagation band is defined as the frequency where the RF transmission of the device is reduced by 3 dB relative to the peak transmission. The band edge is tuned by approximately 50 MHz over 1 mm of displacement of the glass plate, with the tuning rate as high as 100 MHz/mm when the plate is close to the device surface (see Supplementary Fig. S2b). The absolute tuning range and rate reduces with distance as the plate leaves the evanescent field.

# Supplementary Information for

# A Photonic Crystal Receiver for Rydberg Atom-Based Sensing

*Hadi Amarloo, Mohammad Noaman, Su-Peng Yu, Donald Booth, Somayeh Mirzaee, Rajesh Pandiyan, and James P. Shaffer, Quantum Valley Ideas Laboratories, 485 Wes Graham Way, Waterloo, Ontario, Canada N2L 0A7*

In this supplementary section, we provide additional information regarding the details of the off-resonant detection of the RF field enhancement and tuning of the PCR resonance frequency.

**Off-resonant detection:**

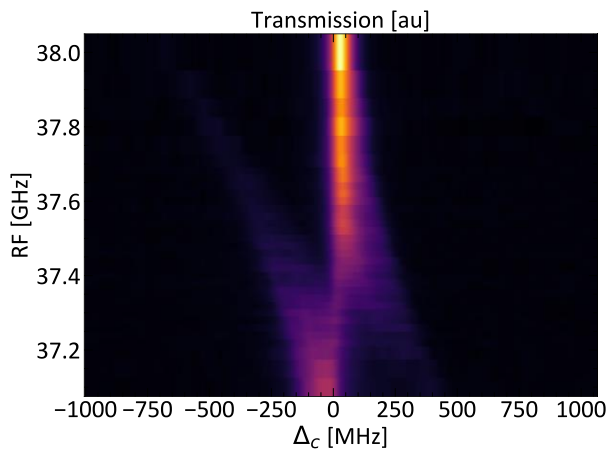

Figure S1: Avoided crossing shows two split peaks and an undeflected central peak. The undeflected peak results from nodes in the RF field in the channel.

Probing the RF field enhancement around the PCR's gain profile provides detailed insight about the PCR's characteristics. RF sensing using Rydberg EIT/EIA utilizes discrete Rydberg states. It is a challenging task to perform a calibration-free measurement of the RF field response over the entire frequency band of the PCR near an atomic transition. We utilized off-resonant, RF field detection and developed methods to perform calibration-free characterization of the PCR. The methods allow for a continuous measurement of the gain of the PCR over a ~1 GHz RF spectral bandwidth. The technique is especially useful in cases where the Stark shifts present in the photonic crystal slot are important, due to Cs sticking to the walls. By analyzing the avoided crossing associated with the RF field interacting with the Cs atoms, the resonance frequency of the PCR can be estimated. Using the avoided crossing is a useful way to obtain the RF field gain. The contour map in Fig. S1 shows the EIT/AT signal as a function of RF frequency inside the PCR. The figure is plotted as a function of detuning of the 509 nm coupling laser from resonance. The measurement is done using elongated probe and coupling beams which capture several nodes/anti-nodes of the RF field inside the PCR. The RF field addressed by the large beam sizes is non-uniform. The measured gain results in a spectrum that is a weighted average of the spatial RF field variations that occur in the PCR. The maximum EIT peak frequency splitting yields the maximum gain of the device. As the RF fields are tuned to different frequencies, the EIT/AT signal frequency splitting shows a peak that is unaffected by the RF field. The unshifted peak results from the presence of RF field nodes along the channel, while the maximally shifted peaks arise from atoms in the antinodes. We fit a 3-Lorentzian peak model to obtain the splitting and hence the RF field. The gain is obtained by comparing the results with a conventional reference vapor cell.

We estimate the amplification by taking the ratio of the Autler-Townes splitting due to the RF field inside the PCR to that of a commercial vapor cell. Our PCR shows a Stark shift from a background DC electric field of ~200 mV/cm, which is accounted for when calculating the amplification. In cases where different

Rydberg states were used for the amplification calculations, the Stark shift corresponding to each of the states is estimated to obtain the gain.

The off-resonance gain is calculated by analyzing the generalized Rabi splitting:

$$\Omega_g^2 = \Omega^2 + \Delta_{RF}^2$$

The data is averaged 300 times to minimize the error.

**Three-Level model:**

To interpret results from the spatial scan along the channel of the PCR, which are complicated by a nearby $47P_{3/2}$ energy level, we implemented a three-level model which includes the effect of the nearby energy level. The Hamiltonian for this system has the form:

$$H = \begin{pmatrix} 0 & \frac{\hbar\Omega_{12}}{2} & \frac{\hbar\Omega_{13}}{2} \\ \frac{\hbar\Omega_{12}^*}{2} & -\hbar\Delta_2 & 0 \\ \frac{\hbar\Omega_{13}^*}{2} & 0 & -\hbar\Delta_3 \end{pmatrix},$$

where $\Omega_{ij}$ is the Rabi frequency of the transition from state $i$ to $j$ and $\Delta_i$ is the detuning of the excitation field to the state $i$. The Rabi frequency is given by the dipole matrix element $\mu_{ij}$ of the atomic transition and the strength of the electric field $E$ driving the transition,

$$\Omega_{ij} = -\frac{\mu_{ij}}{\hbar}E.$$

The detuning is defined as

$$\Delta_i = \omega - \omega_{ij},$$

where $\omega$ is the frequency of the driving field and $\omega_{ij}$ the frequency of the transition.

We can calculate the lines for a given driving field by calculating the eigenvalues of $H/\hbar$. The differences between these eigenvalues are the state splittings, which can be compared to the measured Autler-Townes splitting in the reference cell and the PCR to calculate the driving electric field that matches the measured splitting. The E field in the reference cell is approximately constant with respect to the RF frequency, and the E field in the PCR is influenced by the photonic crystal. We can calculate the power enhancement by taking the ratio of the square of the E field in the PCR and the squared mean of the E field in the reference cell:

$$\text{Power enhancement} = \frac{E_{PCR}^2}{E_{ref}^2}.$$

**Cell resonance tuning:**

Due to manufacturing tolerance, design and fabrication of the PCR can suffer from a mismatch between the device resonance and a target Rydberg state resonance. To facilitate the coincidence of the PCR resonance and that of the atom, the PCR resonance frequency can be tuned by an external glass plate. The tuning can be achieved after the photonic crystal is fabricated. Figure S2 shows the how the PCR tuning works. A glass plate interacts with the evanescent RF field leaking out of the device. The glass plate changes the resonance frequency when it interacts with the RF field. A tuning curve, as the glass plate is moved in the vicinity of the device is shown in Fig. S2 (b).

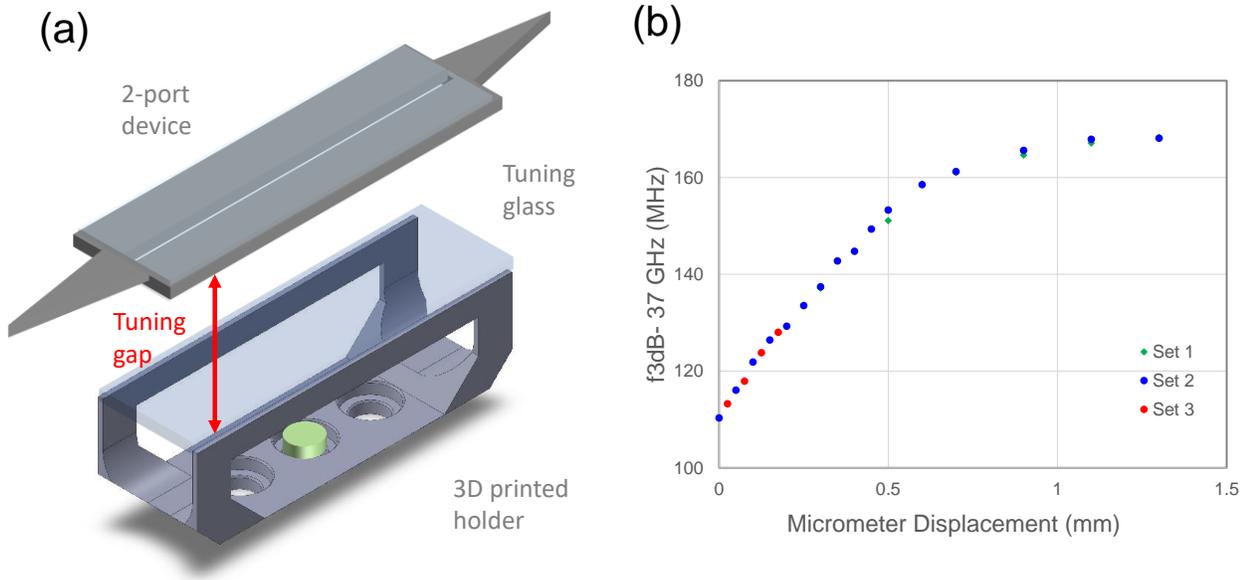

Fig. S2: (a) Illustration of the test tuning setup, where a tuning glass plate is translated to the two-port test device using a plastic holder. The plastic holder and glass plate are mounted on a micrometer-driven positioning stage (not shown). (b) Observed frequency tuning of the band-edge as function of the displacement from the smallest tuning gap allowed by the setup geometry.

**Effect of deviations in hole positions**

The gain variations as a function of frequency in Figure 6 indicate that the impedance matching between the input coupler and the photonic crystal is not ideal. The behavior shown in Fig. 6 can arise from fabrication imperfection, such as deviation of hole positions from the ideal lattice. Photonic crystal hole variation has been shown to impede propagation particularly in slow-light systems.[1] The accuracy of the hole pattern of the photonic crystal can lead to an impedance mismatch. The hole positions have a measured standard deviation of around 20 µm. We performed simulations with this standard deviation of the hole pattern. The simulations show that the RF electric field will build up in the region between the coupler and the photonic crystal as the impedance mismatch can form a cavity. This behavior could also be compared to localization seen in other disordered photonic crystal systems.[2] Figure S3 shows the calculated electric field distribution for the simulation along a line passing though the center of the PCR for two different frequencies; one at the peak of the enhancement shown in Fig. 3, and a point within the band-gap of the PCR device. The RF electric field build-up at the input of the PCR is observed in Fig. S3 in the case where the photonic crystal hole pattern has manufacturing variation. For the frequency in the bandgap, there is no RF electric field in the PCR because there is no coupling into the core of the device. When the incoming RF wave is on resonance with the device, there is coupling into the core of the device.

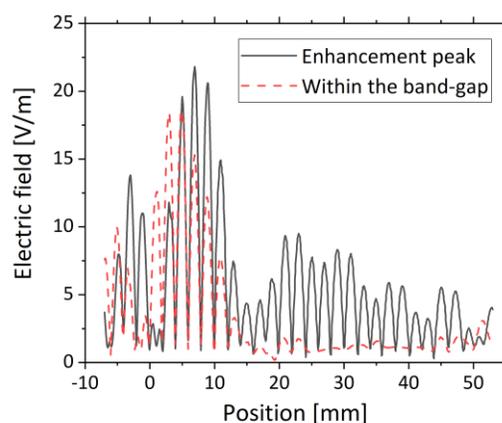

Figure S3: Electric field distribution along a line passing through the center of the PCR at the enhancement peak shown in Fig. 3 (solid black line) and a frequency within the band-gap (dashed red line).